\documentclass[twocolumn]{aastex61}

\usepackage{xspace}
\usepackage{upgreek}
\usepackage{color}
\usepackage{natbib}
\usepackage{amsmath}
\usepackage{booktabs}
\usepackage{multirow}
\usepackage{tabularx} 
\usepackage{footnote}
\usepackage{lipsum}
\usepackage[normalem]{ulem}

\shorttitle{CHIME Pathfinder Search}
\shortauthors{CHIME collaboration}

\begin{document}
\title{Limits on the ultra-bright Fast Radio Burst population from the CHIME Pathfinder}

\correspondingauthor{Liam Connor}
\email{liam.dean.connor@gmail.com}

\author{CHIME Scientific Collaboration} 
\affiliation{The Canadian Hydrogen Intensity Mapping Experiment, DRAO, Kaledan B.C., V0H 1k0}

\author{M.~Amiri}
\affiliation{Department of Physics and Astronomy, the University of British Columbia}

\author{K.~Bandura}
\affiliation{LCSEE, West Virginia University, Morgantown, WV 26505, USA}
\affiliation{Center for Gravitational Waves and Cosmology, West Virginia University, Chestnut Ridge Research Building, Morgantown, WV 26505, USA}

\author{P.~Berger}
\affiliation{Canadian Institute for Theoretical Astrophysics, 60 St. George St., Toronto, ON, M5S 3H8, Canada}
\affiliation{Department of Physics, University of Toronto, 60 St George St, Toronto, ON, M5S 3H4, Canada}

\author{J.~R.~Bond}
\affiliation{Canadian Institute for Advanced Research, Toronto, ON, Canada, M5G 1Z8}
\affiliation{Canadian Institute for Theoretical Astrophysics, 60 St. George St., Toronto, ON, M5S 3H8, Canada}

\author{J.F.~Cliche}
\affiliation{Department of Physics, McGill University, Montreal, Quebec H3A 2T8, Canada}

\author[0000-0002-7587-6352]{L.~Connor}
\affiliation{ASTRON, Netherlands Institute for Radio Astronomy, Postbus 2, 7990 AA Dwingeloo, The Netherlands}
\affiliation{Anton Pannekoek Institute for Astronomy, University of Amsterdam, Science Park 904, 1098 XH Amsterdam, The Netherlands}

\author{M.~Deng}
\affiliation{Department of Physics and Astronomy, the University of British Columbia}

\author{N.~Denman}
\affiliation{Department of Astronomy, the University of Toronto}
\affiliation{Dunlap Institute for Astronomy $\&$ Astrophysics, 50 St. George St., Toronto, ON, M5S 3H4, Canada}

\author{M.~Dobbs}
\affiliation{Department of Physics, McGill University, Montreal, Quebec H3A 2T8, Canada}
\affiliation{Canadian Institute for Advanced Research, Toronto, ON, Canada, M5G 1Z8}

\author{R.S.~Domagalski}
\affiliation{Dunlap Institute for Astronomy $\&$ Astrophysics, 50 St. George St., Toronto, ON, M5S 3H4, Canada}
\affiliation{Department of Physics, University of Toronto, 60 St George St, Toronto, ON, M5S 3H4, Canada}

\author{M.~Fandino}
\affiliation{Department of Physics and Astronomy, the University of British Columbia}

\author{A.J.~Gilbert}
\affiliation{Department of Physics, McGill University, Montreal, Quebec H3A 2T8, Canada}

\author{D.C.~Good}
\affiliation{Department of Physics and Astronomy, the University of British Columbia}

\author{M.~Halpern}
\affiliation{Department of Physics and Astronomy, the University of British Columbia}
\affiliation{Canadian Institute for Advanced Research, Toronto, ON, Canada, M5G 1Z8}

\author{D.~Hanna}
\affiliation{Department of Physics, McGill University, Montreal, Quebec H3A 2T8, Canada}

\author{A.D.~Hincks}
\affiliation{Department of Physics and Astronomy, the University of British Columbia}
\affiliation{Department of Physics, University of Rome ``La Sapienza'', Piazzale Aldo Moro 5, I-00185 Rome, Italy}

\author{G.~Hinshaw}
\affiliation{Department of Physics and Astronomy, the University of British Columbia}
\affiliation{Canadian Institute for Advanced Research, Toronto, ON, Canada, M5G 1Z8}

\author{C.~H\"ofer}
\affiliation{Department of Physics and Astronomy, the University of British Columbia}

\author{G.~Hsyu}
\affiliation{Department of Physics, McGill University, 3600 University St., Montreal, QC H3A 2T8, Canada}

\author{P.~Klages}
\affiliation{Dunlap Institute for Astronomy $\&$ Astrophysics, 50 St. George St., Toronto, ON, M5S 3H4, Canada}
\affiliation{Department of Radiation Oncology, University of Texas Southwestern Medical Center, Dallas, TX 75390}

\author{T.L.~Landecker}
\affiliation{Dominion Radio Astrophysical Observatory, Herzberg Program in Astronomy and Astrophysics, National Research Council Canada}

\author{K.~Masui}
\affiliation{Department of Physics and Astronomy, the University of British Columbia}

\author{J.~Mena-Parra}
\affiliation{Department of Physics, McGill University, 3600 University St., Montreal, QC H3A 2T8, Canada}

\author{L.B.~Newburgh}
\affiliation{Department of Physics, Yale University, New Haven, CT 06520}

\author{N.~Oppermann}
\affiliation{Canadian Institute for Theoretical Astrophysics, 60 St. George St., Toronto, ON, M5S 3H8, Canada}
\affiliation{Dunlap Institute for Astronomy $\&$ Astrophysics, 50 St. George St., Toronto, ON, M5S 3H4, Canada}

\author{U.L.~Pen}
\affiliation{Canadian Institute for Theoretical Astrophysics, 60 St. George St., Toronto, ON, M5S 3H8, Canada}
\affiliation{Canadian Institute for Advanced Research, Toronto, ON, Canada, M5G 1Z8}
\affiliation{Perimeter Institute for Theoretical Physics, Waterloo, ON N2L 2Y5, Canada}
\affiliation{Dunlap Institute for Astronomy $\&$ Astrophysics, 50 St. George St., Toronto, ON, M5S 3H4, Canada}

\author{J.B.~Peterson}
\affiliation{McWilliams Center for Cosmology, Dept. of Physics, Carnegie Mellon University, 5000 Forbes Ave, Pittsburgh, PA 15208, USA}
\affiliation{Dunlap Institute for Astronomy $\&$ Astrophysics, 50 St. George St., Toronto, ON, M5S 3H4, Canada}

\author{T.~Pinsonneault-Marotte}
\affiliation{Department of Physics and Astronomy, the University of British Columbia}

\author{A.~Renard}
\affiliation{Dunlap Institute for Astronomy $\&$ Astrophysics, 50 St. George St., Toronto, ON, M5S 3H4, Canada}

\author{J.R~Shaw}
\affiliation{Department of Physics and Astronomy, the University of British Columbia}

\author{S.R.~Siegel}
\affiliation{Department of Physics, McGill University, Montreal, Quebec H3A 2T8, Canada}

\author{K.~Smith}
\affiliation{Perimeter Institute for Theoretical Physics, Waterloo, ON N2L 2Y5, Canada}

\author{E.~Storer}
\affiliation{Department of Physics, McGill University, Montreal, Quebec H3A 2T8, Canada}

\author{I.~Tretyakov}
\affiliation{Dunlap Institute for Astronomy $\&$ Astrophysics, 50 St. George St., Toronto, ON, M5S 3H4, Canada}
\affiliation{Department of Physics, University of Toronto, 60 St George St, Toronto, ON, M5S 3H4, Canada}

\author{K.~Vanderlinde}
\affiliation{Dunlap Institute for Astronomy $\&$ Astrophysics, 50 St. George St., Toronto, ON, M5S 3H4, Canada}
\affiliation{Department of Physics, University of Toronto, 60 St George St, Toronto, ON, M5S 3H4, Canada}

\author{D.V.~Wiebe}
\affiliation{Department of Physics and Astronomy, the University of British Columbia}

\begin{abstract}

We present results from a new incoherent-beam Fast Radio Burst (FRB) search 
on the Canadian Hydrogen Intensity Mapping Experiment (CHIME) Pathfinder. 
Its large instantaneous 
field of view (FoV) and relative thermal insensitivity allow 
us to probe the ultra-bright tail of the FRB distribution, 
and to test a recent claim 
that this distribution's slope,
$\alpha\equiv-\frac{\partial \log N}{\partial \log S}$, 
is quite small. 
A 256-input incoherent beamformer was deployed on 
the CHIME Pathfinder for this purpose. If the FRB distribution were 
described by a single power-law with $\alpha=0.7$, 
we would expect an FRB detection every few days, 
making this the fastest survey on sky at present.
We collected 1268 hours of data,
amounting to one of the largest exposures of any FRB survey, 
with over 2.4\,$\times$\,10$^5$\,deg$^2$\,hrs. Having seen no 
bursts, we have constrained the rate of extremely bright events
to $<\!13$\,sky$^{-1}$\,day$^{-1}$ above $\sim$\,220$\sqrt{(\tau/\rm ms)}$ Jy\,ms
for $\tau$ between 1.3 and 100\,ms, at 400--800\,MHz. The non-detection also 
allows us to rule out $\alpha\lesssim0.9$ with 95$\%$ confidence, 
after marginalizing over uncertainties in the GBT rate at 700--900\,MHz, 
though we show that for a cosmological population and a large 
dynamic range in flux density, $\alpha$ is brightness-dependent.
Since FRBs now extend to large enough distances that 
non-Euclidean effects are significant, there 
is still expected to be a dearth of faint events and 
relative excess of bright events. Nevertheless we 
have constrained the allowed number of ultra-intense FRBs. 
While this does not have significant implications for
deeper, large-FoV surveys like full CHIME and APERTIF, it does have
important consequences for other wide-field, small dish experiments. 

\end{abstract}

\section{Introduction}

Fast radio bursts (FRBs) are extragalactic, millisecond 
radio transients, 
of which roughly two dozen have been reported 
\citep{2007Sci...318..777L, thornton-2013, 2015MNRAS.447..246P}. 
Though the exact origin of FRBs remains 
elusive, great progress has been made in the last few years alone. 
Uncertainty in their distance scale has decreased by 
twenty orders of magnitude, and the error circle for angular position has 
shrunk by a factor of $\sim$\,billion. A large 
swath of progenitor theories have also been tentatively ruled out 
\citep{2014A&A...562A.137F, 2013ApJ...776L..39K}, 
leaving behind a minority of non-cataclysmic models. 
This came from work establishing 
their extraterrestrial \citep{2015MNRAS.451.3933P} 
and later extragalactic \citep{masui-2015b} nature, as well as the 
discovery that FRB 121102 repeats \citep{2016Natur.531..202S, 2016ApJ...833..177S}.
More recently, \citet{2017arXiv170101098C} were able to 
localize the repeating burst using the VLA, 
leading to the first unambiguous host galaxy identification.
The host was found by \citet{2017ApJ...834L...7T} to be 
a low-metallicity, star-forming dwarf galaxy at $z\approx0.19$.
\citet{2017ApJ...834L...8M} used the European VLBI Network to study 
the radio counterpart, and favor either a low-luminosity AGN 
(discussed in \citealt{2016PhRvD..93b3001R})
or a young neutron star in a supernova remnant 
(proposed by \citet{connor-2016a}; 
developed by \citealt{2016ApJ...824L..32P, 2016MNRAS.461.1498M, 2016arXiv160302891L, 2017arXiv170102370M})
as the progenitor to FRB 121102. 
In the absence of multiple host-galaxy identifications, 
the $\log\!N$--$\log\!S$ test is a useful method of indirectly 
determining the radial distribution of FRBs. 
The volume of the Universe is greater at larger distances, so there tend to be more
faint events than bright ones. Because of this, the sensitive, 
single-dish telescopes that, to date, have discovered all FRBs,
have detected mostly moderate-brightness events due to their 
limited FoV. Therefore, the high-$S$
tail of the FRB distribution has not yet been thoroughly explored.
We parametrize the brightness distribution as a simple 
power-law, such that

\begin{equation}
\mathrm{d}N \propto S^{-(\alpha+1)} \mathrm{d}S,
\end{equation}

\noindent where $S$ is flux density and $N$ is 
number of events. When integrated, this gives 
$N(>S) \propto S^{-\alpha}$, which we refer to as the 
brightness distribution.
This one-parametric class of 
models has a single special value, $\alpha=3/2$, 
corresponding to a non-evolving population of 
sources in a Euclidean spacetime. 
It is worth pointing out that this value is not 
limited to standard candles, so long as there is no statistical 
relationship between distance and luminosity or volume density. 
The 3/2 case also holds for 
anything proportional to flux density, so fluence or 
signal-to-noise can be used in place of $S$.

Several groups have tackled the $\log\!N$--$\log\!S$ problem.
\citet{vedantham2016} argued that a surplus of multi-beam 
detections at Parkes implied a comparatively flat fluence distribution, 
with $0.52<\alpha<1.0$. 
\citet{Oppermann16} used the ratio of observed signal-to-noise, $s$, 
to the search threshold, $s_{\rm min}$, to test the Euclidean hypothesis, 
motivated by the fact that it is model independent and does not 
suffer from survey incompleteness.
This essentially reinstituted the classic $\left < V/V_{\rm max}\right >$-test
that was used to show the cosmological nature of quasars \citep{schmidt-1968} 
and gamma-ray bursts (GRBs; \citealt{mao92}). They found consistency with a Euclidean 
distribution, but neither the Lorimer burst \citep{2007Sci...318..777L} nor FRB 150807 
\citep[then-unpublished]{2016Sci...354.1249R}
were included in their analysis, whereas \citet{vedantham2016} used 
both. \citet{ata2011} carried out a wide-field FRB search in 
``Fly-Eye'' mode on the Allen Telescope Array (ATA) with roughly one 
quarter the exposure of our Pathfinder survey, though too few FRBs
had been observed at the time to put limits on $\log\!N$--$\log\!S$. 
\citet{vedantham2016} concluded separately that ATA's non-detection 
rules out $\alpha\lesssim0.6$. \citet{2015MNRAS.451.3278M} argued 
that the apparent deficit of events at low Galactic latitudes may be explained
by a steep brightness distribution, with $\alpha>2.5$. However, such steep 
$\log\!N$--$\log\!S$ are now disfavored by the data.

If \citet{vedantham2016} are correct
and the brightness distribution is much flatter than expected, 
then the implications for survey design are striking.
They point out that for $\alpha<1$, 
small dishes are actually preferred to large dishes
because the high number of bright events favors sky coverage 
over sensitivity. Survey speed, $\Gamma$, which we take to be the rate 
at which a given experiment detects FRBs,
is given by the product of 
field of view and a thermal sensitivity term 
raised to the power of $\alpha$. Sensitivity increases with collecting 
area, which scales quadratically with dish diameter, $D$, and 
beamsize goes as $1/D^2$. Therefore, 

\begin{align}
	\Gamma &\propto \mathrm{FoV} \times \mathrm{sensitivity}^\alpha \nonumber \\
		   &\propto D^{2(\alpha-1)},
\end{align}

\noindent and survey speed decreases with increasing dish size for
flat distributions  ($\alpha < 1$). Using an incoherent-beam search on the 
pre-existing Canadian Hydrogen Intensity Mapping Experiment (CHIME) 
Pathfinder, we are able to test the low-$\alpha$ hypothesis with limited 
time on sky, based on similar arguments. The incoherent beam 
is generated by adding up the signals from all antennas after squaring 
their voltage time streams, erasing relative phase information. This 
produces a less sensitive beam than the coherent case, for which 
phase is preserved, but is the size of the full primary beam. 

\begin{figure*}
  \centering
\includegraphics[trim={0in, 0.5in, 0in, 0in}, height=5in, width=0.9\textwidth]{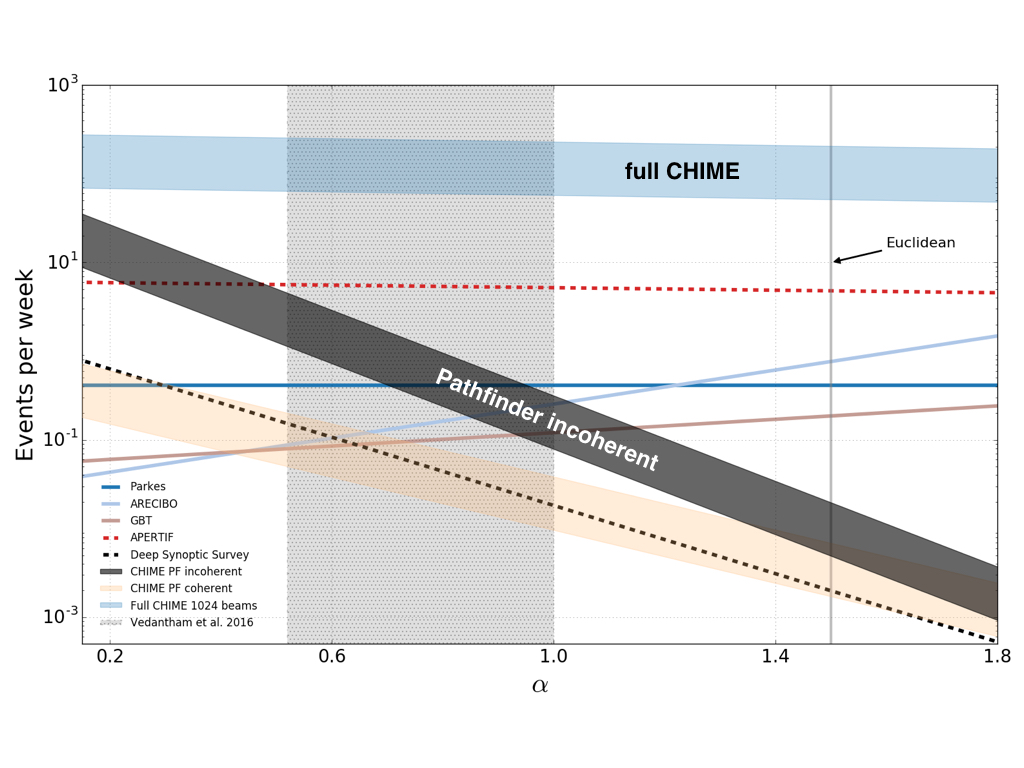}
\caption{Weekly event rates for several different surveys 
plotted as a function of brightness distribution parameter 
$\alpha$, pinned to the Parkes rate (hence its flatness), 
similar to what is done in Eq.~\ref{eq-mu} with GBT.
The three solid lines are surveys that have 
already found FRBs. The light grey dotted region shows the 90$\%$
confidence interval of $\alpha$ proposed by \citet{vedantham2016}.
The thick colored regions allow for uncertainty in the rate 
below 1.4\,GHz, and the dashed curves are for 1.4\,GHz surveys still in the 
commissioning phase. Interestingly, the modestly sized  
CHIME Pathfinder in incoherent mode (dark grey)  
would be the fastest survey to date in incoherent mode, if $\alpha\lesssim0.75$. This is 
due to its 200\,deg$^2$ primary beam and the importance of FoV for 
small $\alpha$.}
\label{fig-event-rates}
\end{figure*}

We expect a coherent beam from $N_a$ dual-polarization antennas will be $\sqrt{N_a}$ 
times more sensitive than an incoherent beam 
from the same set of inputs, assuming noise is mostly uncorrelated between 
receivers. 
The factor of $N_a$ in incoherent beam 
solid angle ultimately wins though---dramatically so for small $\alpha$, 
as can be seen by comparing the dark grey and orange regions in 
Fig.~\ref{fig-event-rates}.
If we take the ratio of the incoherent survey speed to 
coherent survey speed, assuming equal bandwidth and signal-to-noise 
cut-off, we get,

\begin{align}
\frac{\Gamma_{\rm inc}}{\Gamma_{\rm coh}} &= \frac{N_a \Omega_i}{\Omega_i} 
\times \left ( \frac{\sqrt{N_a} \,G_i/T_{\rm sys}}{N_a\,G_i / T_{\rm sys}} \right )^\alpha \nonumber \\
&= N_a^{1-\alpha/2}, 
\label{eq-inc-coh}
\end{align}

\noindent where $G_i$ and $\Omega_i$ are the gain and beam solid angle
of a single feed. The Pathfinder, 
for which $N_a=128$, should benefit from a factor of about 23 in speed-up 
for $\alpha=0.7$ when going from a coherent to an incoherent beam.

The ``full'' CHIME FRB project is expected to see multiple events per day, 
making it the fastest survey on sky \citep{connorgbt, 2017arXiv170107457C}. 
This is mainly due to its ability to search all $\sim$\,10$^3$ 
coherently-formed beams, 
with near 100$\%$ duty-cycle, filling its full $\sim$\,200\,deg$^2$ FoV
primary beam at all times \citep{cherry2017}.
Because full CHIME also has appreciable collecting area (8000\,m$^2$), 
it is relatively $\alpha$-independent, which is also the case for 
fast upcoming surveys like APERTIF \citep{leeu14} and UTMOST \citep{2016MNRAS.458..718C}.
This is not true for the CHIME Pathinder, which has a similar 
design to full CHIME, but less collecting area 
and its beam-forming backend is presently capable of 
processing only one synthesized full-polarization beam.
This can be seen in 
Fig.~\ref{fig-event-rates}, where we plot the expected number 
of detected FRBs per week as a function of $\alpha$, 
both for existing experiments and those in the commissioning stage. 
Large-FoV, highly-sensitive instruments like CHIME (light blue solid region)
and APERTIF (dashed red curve; \citealt{leeu14}) are able to see faint events, 
as well as the rarer bright events. However, specialized instruments like 
the incoherent-beam CHIME Pathfinder (dark grey solid region) and
the Deep Synoptic Array\footnote{www.astro.caltech.edu/$\sim$srk/Workshop/BnE2016\_NB.pdf}
(dashed black curve) are only competitive 
if $\alpha$ is small. Moderate FoV instruments like the Parkes 
Multibeam Receiver and Arecibo's ALFA are orders of magnitude faster 
than the incoherent Pathfinder search for the Euclidean case, 
but several times slower if $\alpha < 0.8$.

\pdfpageattr {/Group << /S /Transparency /I true /CS /DeviceRGB>>}


In this paper we discuss the new incoherent-beam FRB survey 
on the CHIME Pathfinder. Its development was motivated by two points.
Given its large instantaneous FoV but poor sensitivity,
we could very quickly test the low-$\alpha$ hypothesis. 
And if $\alpha$ really were significantly smaller than 
3/2, we would have set up---with little cost--- 
a survey faster than the Parkes Multi Beam. We outline 
this experiment in Sect.~\ref{sec-pathfinder}, 
including the development of its beamforming and tree-dedispersion 
pipelines. 
In Sect.~\ref{sec-results} we discuss our non-detection in 
$\sim$\,53 days of data and the constraints on $\alpha$. 
We then go over the implications for other similar surveys, 
and discuss various astrophysical reasons for our non-detection
in Sect.~\ref{sect-discussion}.

\section{CHIME Pathfinder}
\label{sec-pathfinder}

A pathfinder instrument for CHIME was constructed 
at the Dominion Radio Astrophysical Observatory (DRAO)
in Penticton, British Columbia, was brought online 
in late 2013 \citep{2014SPIE.9145E..22B}.  
Its purpose is to act as both a proof-of-concept instrument 
and a debugging tool for the full CHIME, whose 
highly ambitious primary science 
goal of 21\,cm intensity mapping requires 
considerable precision in calibration.
The Pathfinder consists of two north-south 37\,m-long, 20\,m-wide 
cylindrical mesh reflectors, whose focal lines are each instrumented 
with 64 linear dual-polarization antennas for a total of 256 inputs.
This is roughly an order of magnitude smaller in scale than 
full CHIME, which has a total of 2048 inputs on four 
100\,m-long, 20\,m-wide reflectors. More information 
about CHIME's pathfinder instrument can be found in \citep{2014SPIE.9145E..22B, 
2014SPIE.9145E..4VN, 2015PhRvD..91h3514S, berger16}.

The stationary, cylindrical reflector design makes CHIME 
a wide-field transit telescope. Since its dishes are
aligned north-south, it only focuses light 
in the east-west direction, resulting in a primary beam 
that spans $\sim$\,150$^\circ$ in declination and 1--2$^\circ$ in hour-angle. 
North-south spatial resolution is recovered either by beamforming or 
computing the full $N^2$-correlation matrix, both of which 
are done in the Pathfinder's correlator (for more 
details, see \citealt{2015arXiv150306189R}).

\subsection{Beamformer}

Since late 2015, we have had a working beamforming back-end
in the Pathfinder. The beamformer is an OpenCL kernel 
run on a 16-node GPU cluster, which is the X-engine 
of the Pathfinder's hybrid FX-correlator \citep{2015arXiv150306202D}.
It is run in a commensal mode with the more computationally intensive 
$N^2$-correlation that is used for the cosmology experiment.   
Initially, the beamformer produced a single coherent tracking beam 
that was used for pulsar observations and a 
preliminary FRB search. Once it was realized that an 
incoherent beam may potentially provide an enormous increase 
in search speed, the coherent beamforming kernel was modified 
to first square, then sum, incoming voltages. 
Channelized data arrive at each of the 16 GPU nodes 
from the custom F-engine electronics as 4-bit real, 
4-bit imaginary offset encoded integers.
Once these voltages are squared and 
summed across the array, they are reduced to 8-bit unsigned integers.
Signals from all 256-inputs, but only one sixteenth of the frequencies, 
are processed on each node. The beamformed 
data are sent to a separate acquisition node over a 10 Gigabit Ethernet
at 6.4\,Gbps in the
VDIF specification\footnote{www.vlbi.org/vdif/docs/VDIF\_specification\_Release\_1.1.1.pdf}. 

\begin{figure*}
\includegraphics[trim={.5in, 0in, 0.5in, 0in}, width=\textwidth]{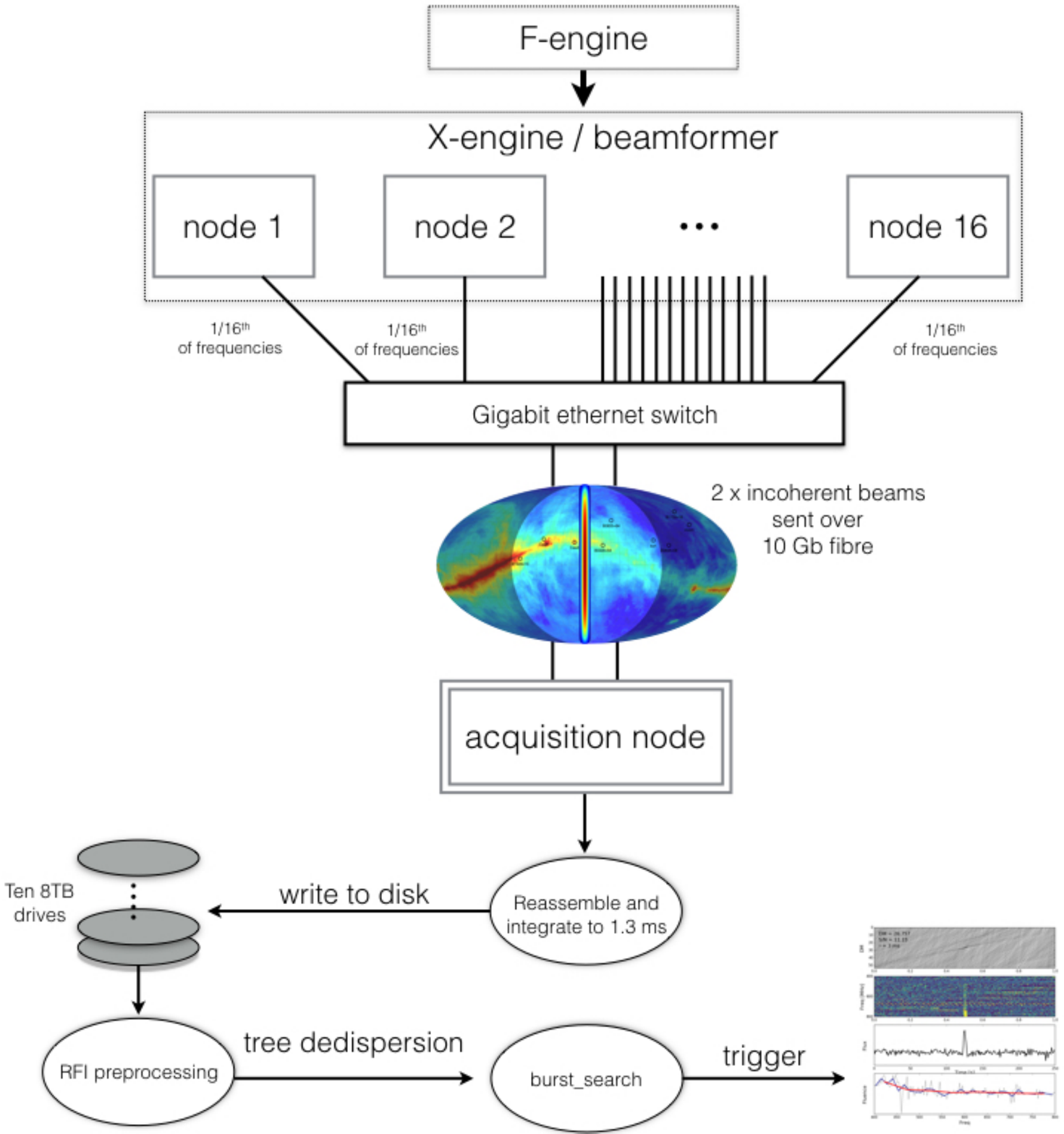}
	\vspace{0.5 cm}
	\caption{Block diagram outlining the Pathfinder's incoherent-beam FRB search. Data are 
	channelized in a 16-FPGA F-engine, sent to the beamforming 
	kernel in a GPU-based X-engine, then sent to an acquisition node 
	and written to disk. These 
	intensities are searched offline using a tree-dedispersion code, 
	after RFI preprocessing.}
\label{fig-block-diagram}
\end{figure*}

For the incoherent beamforming kernel, the intensities
arrive at the acquisition node at full time and frequency 
resolution, 2.56\,$\mu$s and 390.625\,kHz respectively. 
The squared and summed signals from our two orthogonal polarizations 
arrive separately, and are not summed until further down the pipeline. 
Since different frequencies arrive from different nodes, packets arrive out 
of order and must be unscrambled. A real-time, multi-threaded acquisition code was developed to handle 
this\footnote{https://github.com/kmsmith137/ch\_vdif\_assembler}. 
It writes to disk either assembled voltages (in the coherent beamforming case) 
or intensities integrated to 1.3\,ms (in the incoherent case).
The latter are written to HDF5 files for offline processing.
A diagram of the incoherent back-end and search pipeline 
is shown in Fig.~\ref{fig-block-diagram}. The custom data processing pipeline
was built specifically for this experiment. 

\subsection{FRB Search}

We run a modified tree dedispersion algorithm on the data to 
search for FRBs with dispersion measures (DMs) between 20 and 2000\,pc\,cm$^{-3}$ 
and widths between 1.3 and 100\,ms. 
The package {\tt burst\_search}\footnote{https://github.com/kiyo-masui/burst\_search} 
was first developed to search GUPPI data from the Green Bank Telescope, 
and successfully found FRB 110523 \citep{masui-2015b}. We modified 
this code to search both real-time Pathfinder data streams 
and offline, integrated intensity data. The data were 
broken up into 80-second total-intensity (Stokes I) 
arrays that overlapped by 15\,s 
with the previous block. We do not search over spectral index.
If the largest S/N value in a given block exceeded our threshold 
of 10, a trigger was written to disk, along with plots of the 
event. The large number of low-DM triggers jeopardized our seeing 
extragalactic events, so we searched ranges 20--200\,pc\,cm$^{-3}$, 
200--525\,pc\,cm$^{-3}$, and 525--2000\,pc\,cm$^{-3}$ separately.

\subsubsection{RFI}
\label{sect-rfi}

Due to the incoherent beam's sensitivity to the horizon, 
radio frequency interference (RFI) was a significant concern.
When we searched raw data without any masking or de-trending, 
an event above the S/N threshold would occur in every processed 
block of data. A large fraction of these were due to the recently-introduced 
Long Term Evolution (LTE) wireless communication band 
around 700\,MHz, which fluctuates on millisecond time-scales. By masking 
these and other persistent RFI frequency channels, most false positives 
could be avoided. But the data still needed preprocessing, so a series 
of filters was applied to each block of intensities before the tree-dedispersion 
search was run. 
This included a 6\,$\sigma$ outlier cut in frequency, which removes bad channels.
Bandpass calibration is done by dividing our data by the time-averaged 
DC power within an 80-s block. We then apply a highpass filter 
that used a 100\,ms Blackman window function, which sets our maximum 
search width.  
The effectiveness of this RFI-preprocessing was verified 
using transits from pulsar B0329+54, as well as simulations of FRB events;
after preprocessing these events would be detected, 
but without the filtering, B0329+54 and injected events would go 
undetected due to strong RFI occurring during the dispersed pulse.
After pre-processing we reduced our false-positive 
rate to roughly one per 30 minutes.

\subsection{Survey parameters}

In order to estimate an expected FRB rate we need to know 
the telescope's sensitivity, beamsize, and the effects 
of dispersion smearing, which can be difficult to determine. 
For example, 
the reduction in search speed of dispersion smearing is calculable only 
if the DM and width distributions of ultra-bright FRBs are known. 
Below we provide estimates of each of these quantities and 
their associated uncertainty.


\subsubsection{Beamsize}

Extensive work has been done both to simulate and map the primary beams of each 
Pathfinder antenna. Using the east-west holographic measurements 
made by \citet{berger16}, and simulations of the 
north-south beam using the reflector antenna software 
{\tt GRASP}\footnote{http://www.ticra.com/products/software/grasp},
we adopt half-power beam solid angles of 270, 225, 
155, and 110 square degrees at 430, 525, 625, and 750 MHz, 
respectively. For most of our analysis 
we take their mean, $\sim$\,190\,deg$^2$, to be our beamsize. 
However, it seems likely that this is a conservative estimate; early holography 
measurements indicate that the true north-south beam on-sky  
is larger than the beam produced in simulation. 

\subsubsection{Sensitivity}
\label{sect-sensitivity}

We were able to test our expected sensitivity using two 
distinct methods. The first method uses the fractional power increase 
from the transits of bright point-sources such as Cassiopeia A, Cygnus A, 
and Taurus A (Crab nebula), to estimate the baseline $T_{\rm sys}$. This ``Y-factor method"
measures the average system temperature of \textit{the individual antennas}, 
but does not measure the effective $T_{\rm sys}$ of our incoherent beam. 
The point-source transits give what we expect, 
namely an average system temperature per antenna of $\sim60$\,K assuming an 
aperture efficiency of 50$\%$ \citep{Davis_2012}. 
The second method, which is more relevant to 
our search, comes from using the radiometer equation 
with measurements of single pulses of B0329+54, and indicates 
higher-than-expected noise. The difference between the two methods is that 
the latter measures an actual RMS of the final incoherent beam, 
so it probes the way the noise averages down after we sum across the array.

B0329+54 is the brightest visible pulsar in our band, 
and fortuitously is only $\sim$\,5 degrees off-zenith at our 
latitude.
According to 
the Australia Telescope National Facility (ATNF) Pulsar Database, 
B0329+54 has flux densities at 400\,MHz and 1.4\,GHz of
$S_{400}^{\nu} = 1500$\,mJy and $S_{1400}^{\nu} = 200$\,mJy, respectively \citep{atnf}.
With a pulse width of 6.6\,ms and a 714\,ms period, 
its interpolated flux density at 600\,MHz when 
it is ``on'' is 84\,Jy, assuming a power-law index $\gamma=1.61$. 
To calculate the expected S/N from a dedispersed, frequency-averaged
time-series, the average flux, $\left < S \right >_\nu$, can be 
estimated by summing in quadrature. This gives an effective 
flux density of $\sim$\,96\,Jy. If we then measure the average S/N 
of single B0329+54 pulses, we can use $\left < S \right >_\nu$
to constrain the system's sensitivity. This is done 
with the radiometer equation,

\begin{equation}
\mathrm{S/N} = \frac{\left < S \right >_\nu \sqrt{2 B_{\mathrm{eff}} \tau}}{S_{\rm sys}},
\label{eq-radiometer}
\end{equation}

\noindent where $B_{\rm eff}$ is the effective bandwidth, $\tau$ is the pulse duration, 
and the factor of 2 is the number of polarizations. 
$S_{\rm sys}$ is the system-equivalent flux density (SEFD), 
which is simply the ratio of system temperature to forward gain,
$T_{\rm sys}/G$. We opt for the SEFD since $T_{\rm sys}$ and $G$
are often degenerate, and we do not need to distinguish between 
the two for our purposes. 

From the several dozen B0329+54 transits we have in our dataset, 
over 50,000 individual pulses were observed. Analyzing the stored data, 
we find at beam center, 
the mean S/N was $\sim$10, so $S_{\rm sys} = 2 \times 10^4$\,Jy, 
using $B_{\rm eff}$ and $\tau$ from Table ~\ref{tab-parameters}.
This is a few times larger than expected, 
which seems to be caused by excess noise on time-scales 
$\lesssim20$\,ms, leading to larger RMS on all time-scales. 
This excess is also seen in the noise power spectrum 
at high temporal frequencies, and may be caused by intermittent RFI. 
The discrepancy between the Y-factor method and the S/N of B0329+54 pulses 
would then come from correlated RFI-induced noise not beating down as $\sqrt{N_a}$ 
as we sum all antennas in the beamformer.

We also collected roughly 10$^3$ Crab giant pulses (GPs). Due to their 
steep brightness distribution and uncertainty in absolute flux density, 
they were not directly used as a calibrator. However, by comparing to 
a large set of GPs observed in our band with the Algonquin Radio Telescope 
(Main et al., in prep), we find our rate of 2--8 GPs
per minute with $s>10$\,$\sigma$ to be consistent with the brightness 
distribution they found.


\begin{table}
\centering
\caption{Parameters of the CHIME Pathfinder and its
		incoherent FRB survey. $N_a$ is the number of dual-polarization antennas,  
		which is half the number of beamformer inputs. 
		}
\label{tab-parameters}
\begin{tabular*}{\columnwidth}{c @{\extracolsep{\fill}} lll} \toprule
        			  & PARAMETER										& VALUE 		\\[0.1mm]\midrule
				      & $N_a$                   						& 128           \\[1mm]
\textbf{CHIME Pathfinder}          & $A_{\rm geo}$ {[}m$^2${]} 			& 880           \\[1mm] 
                      & freq {[}MHz{]}          						& 400--800       \\[1mm]
                      & $N_{\rm chan}$      							& 1024          \\[5mm]
					  & $S_{\rm sys}$ {[}Jy{]}  						& $2\times10^4$ \\[1mm]
                      & $T_{\rm obs}$ {[}hrs{]} 						& 1268          \\[1mm]
\multirow{2}{*}{\begin{minipage}{0.5in}\textbf{Incoherent FRB search}\end{minipage}}                       & $\Omega$ [deg$^2$]							    & 190           \\[1mm]
                      & $B_{\rm eff}$ [MHz]								& 312           \\[1mm]
                      & $s_{\rm min}$ [$\sigma$]  									& 10            \\[1mm]
                      & $\mathcal{S}_{\rm min}^{3\rm\,ms}$ [Jy]  		& 125           \\[1mm]
                      & Exposure [deg$^2$ hrs]                       	& 241,000       \\
\end{tabular*}
\end{table}

\subsubsection{DM Smearing}
\label{sect-smear}

Though full CHIME will ``upchannelize" its data (increase the frequency resolution 
after the initial channelization step) \citep{cherry2017}, 
the incoherent Pathfinder search was carried out with 
the nominal 1024 channels at 390-kHz resolution. This leads to ``DM smearing'' for 
highly dispersed events, which broadens 
the pulse and reduces S/N. If the FRB's intrinsic width 
is $t_i$, is scattered to a width $\tau$, and is sampled at $t_{\rm samp}$,
the minimum flux density to which we are sensitive is 
increased as,

\begin{equation}
S'_{\textrm{min}} \rightarrow S_{\textrm{min}} \times 
\left ( \frac{t_I}{\sqrt{t_{\rm samp}^2+\tau^2 + t^2_i}} \right )^{1/2},
\label{eqn:sens_smear}
\end{equation}

\noindent where $t_I$ is the final pulse width \citep{2014ApJ...792...19B}.
Using $\Delta \nu$ as the frequency 
resolution, and $\nu_c$ as the central frequency, 
the effective pulse width can be calculated by 
adding in quadrature the other broadening elements.
This is done as follows,


\begin{equation}
t^2_I = \tau^2 + t^2_i + t^2_{\textrm{samp}} + t^2_{\textrm{DM}}
\label{eqn:smearing}
\end{equation}

\noindent where 

\begin{equation}
t_{\textrm{DM}} = 8.3 
\left ( \frac{\rm DM}{\rm\,pc\,cm^3} \right)
\left ( \frac{\Delta \nu}{\rm1\,MHz} \right)
\left ( \frac{\nu_c}{\rm1\,GHz} \right)\, \mu\rm s.
\end{equation}

In this survey the smearing term will dominate 
the sampling time and, probably, intrinsic width,
for high DMs. Scattering is less constrained. 
For example, a burst with DM=776\,pc\,cm$^{-3}$ (the median 
DM on FRBcat \citealt{petrofffrbcat})
that was intrinsically 1\,ms, sampled at 1.3\,ms, and scattered to 
5\,ms (roughly the case for GBT FRB 110523 if it were 
observed at 600\,MHz), would be $\sim$12\,ms
in duration if observed on the Pathfinder. Therefore, if all FRBs
had the parameters of that hypothetical 
burst, Eq.~\ref{eqn:sens_smear} tells us that 
the current Pathfinder search would be $\sim$\,$6^{\alpha/2}$ times 
slower than a sufficiently upchannelized Pathfinder search. But not all FRBs will have those exact parameters, and the number of degrees of freedom 
in Eq.~\ref{eqn:smearing} makes predicting the effects of 
smearing for high-DM events difficult. 

Fortunately, there is reason to think this would not be a major issue. 
The only way the incoherent-beam Pathfinder search will see anything 
is if $\alpha$ really is small (see the low detection rate 
for $\alpha>1.2$ in Fig.~\ref{fig-event-rates}).
That would mean the IGM is doing a significant fraction of 
the dispersion, 
in which case brightness anti-correlates with DM, as 
nearby sources have less intervening plasma. In other words, our survey 
only probes the ultra-bright, nearby subset of the FRB population, 
and their low DMs will not greatly reduce the survey's sensitivity. Indeed, the two 
sources whose inferred flux density was orders-of-magnitude greater than 
the FRB median (the Lorimer burst and FRB 150807) both had extragalactic DMs
less than 350\,pc\,cm$^{-3}$ \citep{2007Sci...318..777L, 2016Sci...354.1249R}. 

\section{Results}
\label{sec-results}

In 1268 hours of data several thousand triggers were produced 
with signal-to-noise greater than 10. Each was inspected by eye,
and almost every ``event'' was discernibly non-astronomical. For example,
the incoherent beam's susceptibility to RFI means that
most triggers were narrow-band or had unusual discontinuities 
in their frequency-collapsed profile. As discussed in Sect.~\ref{sect-rfi}, 
some of these false-positives were caused by strong interference 
flickering on time-scales of tens of milliseconds.
The handful of marginal events were analyzed further, but no FRBs were found. Having 
seen zero events, we can ask how unlikely that outcome was 
and therefore put a lower limit on $\alpha$. But first, 
we must verify that if there were an FRB in our beam, 
we would have detected it.

\subsection{Completeness}

There are many ways for a transient search 
to not see something, so care must be taken in 
verifying a survey's completeness and reliability. 
Since the incoherent beamformer does not do any spatial
filtering, we see the whole northern sky 
each day with CHIME's large north-south primary beam. 
Giant pulses (GPs) from the Crab and single pulses from B0329+54 were used 
to ensure the search pipeline was working and 
that each day's data were good. All other pulsars---including 
GP-emitting sources like B1937+21---are too faint to see individual pulses. 
This is also true for known RRATs. 
For the two sources we could see, we 
found that, even when the sources were entering the beam and 
the maximum S/N in a given 80-s block was around our 
search's threshold, $s_{\rm min}$, 
individual pulses still triggered and 
were easily recognizable as pulses. 
We detected B0329+54 pulses and Crab GPs in all 
of their respective transits during our observing campaign.
However, getting a fractional completeness---the ratio of the number of detections 
to number of events---is difficult for these sources.  
This is due to their pulse-to-pulse intensity fluctuations, which cause their 
S/N to fall below our threshold, and the fact that we trigger 
only on the brightest event above 10\,$\sigma$ in each 80-s block of data.
In order to quantify our fractional completeness, we injected 
signals into our data with $\mathrm{DM}=400$\,pc\,cm$^{-3}$
at a range of brightnesses. We find that for the injected signals
whose expected resultant S/N$\gtrsim15$, effectively all pulses are recovered
and our completeness is above 99$\%$.
For events whose recovered S/N 
is between 10--12, we detect $\sim$90$\%$ of the simulated bursts. Fortunately, as we 
show in Fig.~\ref{fig-triggers}, for $N(>S)$ power-laws with $\alpha<1.5$, most
FRBs are expected to be detected above 15\,$\sigma$.

\begin{figure*}
  \centering
\includegraphics[trim={0in, 0in, 0in, 0in}, width=0.925\textwidth]{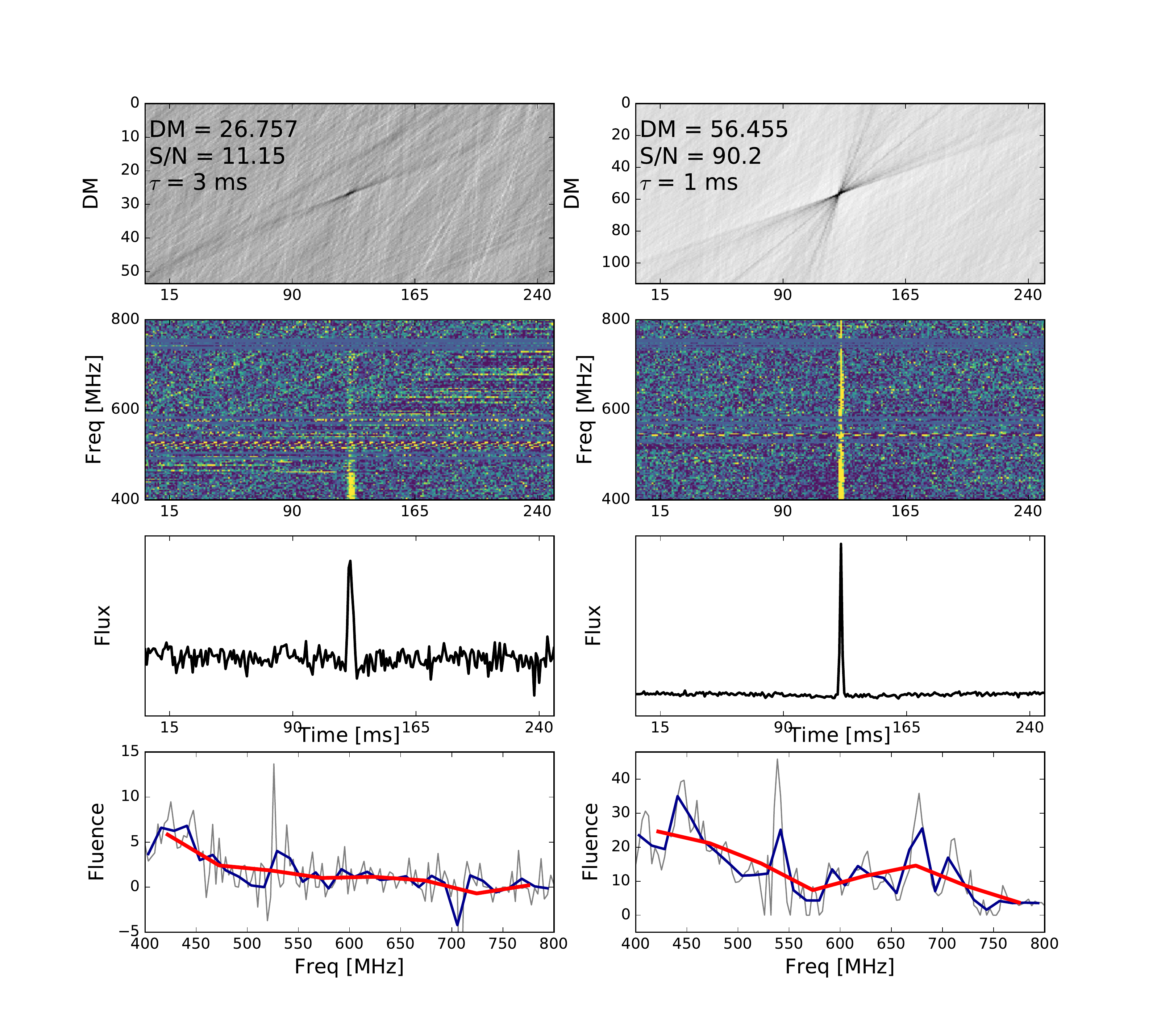}
   \caption{Example triggers from our {\tt burst\_search} pipeline. Each ``event'' 
    produces a figure with four panels. From top to bottom: amplitude in 
    DM vs. arrival time; the 
    dedispersed intensity in frequency and time; the frequency-collapsed pulse profile; 
    and a fluence spectrum of the pulse for three different binnings. 
    The left 
	figure shows a typical single pulse from B0329+54 with $s\sim11$. 
	Even when events were right around $s_{\rm min}$, 
	Crab GPs and B0329+54 pulses were unambiguously celestial, in the sense that 
	they were broad-band, isolated in DM vs. arrival time, and were
	detected as single peaks in the frequency-collapsed pulse profile. 
	This is promising given none of the unexpected triggers had this property.
	The right figure shows a high-significance Crab GP, with $s\sim90$.
	This established that our RFI-preprocessing should not remove ultra-bright events,
	even in the presence of significant frequency structure.}
   \label{fig-example_trigger}
\vspace{1cm}
\end{figure*}

\begin{figure}
  \centering
  \vspace{1 cm}
   \includegraphics[trim={.3in, .2in, .2in, 0in}, width=0.48\textwidth]{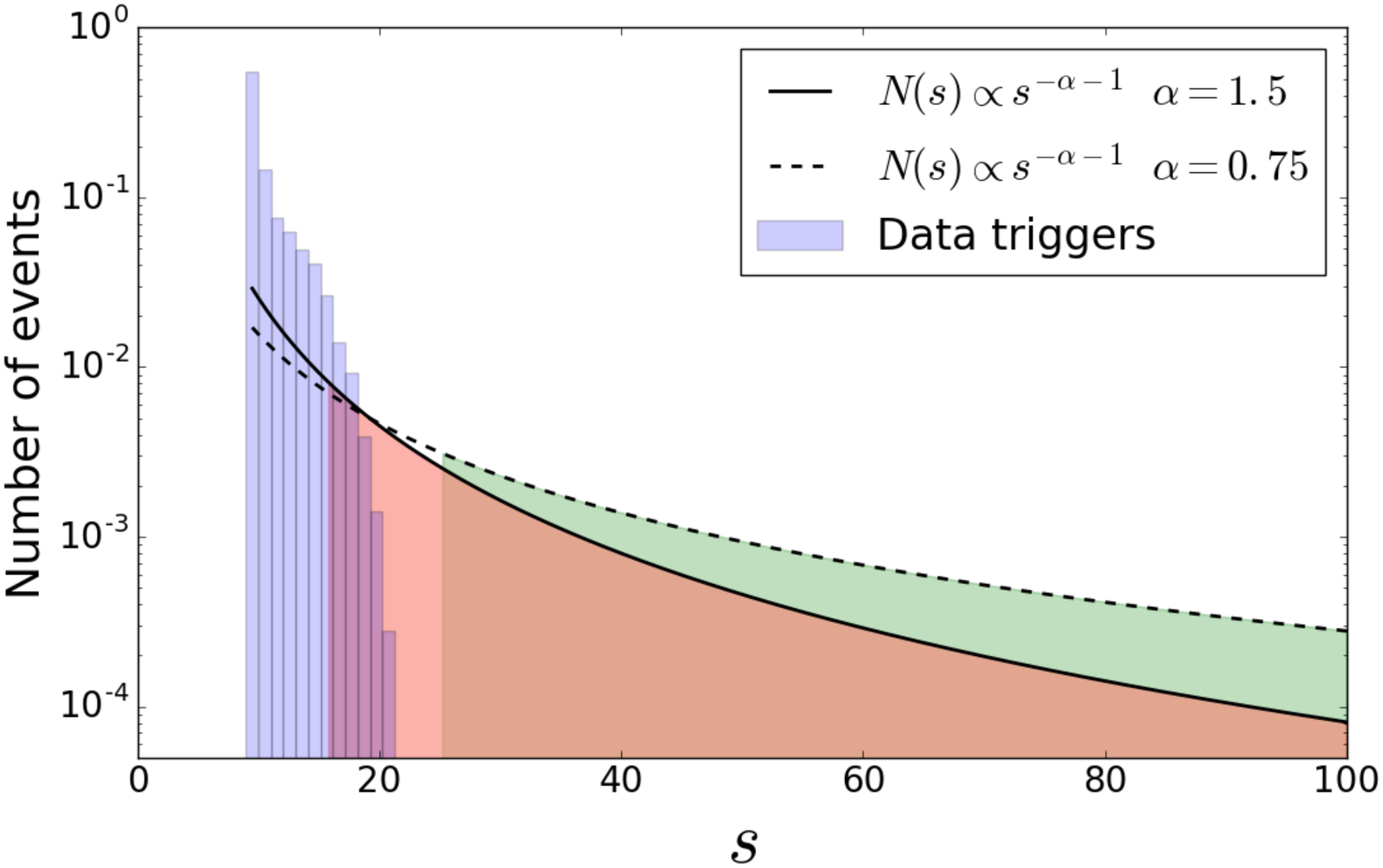}
   \caption{A comparison between the S/N distribution of false positive 
   triggers found in the data with the distribution expected for FRBs. 
   The purple histogram shows the normalized S/N counts for 3470 triggers in the data, each inspected by eye. The solid 
   and dashed curves show the expected power-laws with $\alpha=1.5$ (Euclidean) and
   $\alpha=0.75$, respectively. The shaded region under each curve represents half of the 
   probability mass, meaning for the values of $\alpha$ that this survey can probe, 
   50$\%$ of events should have higher S/N than any false-positive in
   the dataset.}
   \label{fig-triggers}
\end{figure}

Two examples of the output 
of our search are seen in Fig.~\ref{fig-example_trigger}, which show 
four different visualizations of the data for each event. The top panel of 
each trigger plot shows the burst's amplitude in DM / arrival time space. 
The second panel from the top shows a frequency / time 
intensity array after dedispersing the pulse to 
the maximum likelihood DM. The next panel shows 
a frequency-averaged pulse profile, and the final 
panel shows fluence plotted against frequency for 
three different binnings. The B0329+54 
trigger illustrates that pulses near the cut-off are still 
easily identifiable. The Crab trigger shows that high-S/N events 
are not excised by our RFI-preprocessing.



The clarity of B0329+54 pulses or Crab GPs close to 10\,$\sigma$ 
is in stark contrast to the vast majority of unexpected triggers. 
Several thousand events were inspected 
by eye, almost all of which were unequivocally false positives.
They would have power only in a few frequency channels, or 
would look like step-functions in time.
The borderline events were followed up by analyzing directly 
the data around the event, but ultimately there 
were no triggers that looked like broad-band, single-DM pulses. 
The triggers produced tended to be very low DM, which is why 
we partitioned the full DM range into three groups. RFI triggers 
cluster around the minimum search DM as well as the signal-to-noise 
threshold. The latter effect is shown in Fig.~\ref{fig-triggers}. The 
light purple histogram is the S/N distribution of 3470 triggers. 
Almost half of them had $10\leq s\leq11$, whereas only 7$\%$ and 13$\%$ 
of FRBs would be within 1\,$\sigma$ of the threshold, assuming 
$\alpha=0.75$ and $\alpha=1.5$ respectively.

\subsection{Constraints on $\alpha$}

If we treat the arrival times of detectable FRBs as Poissonian, 
we can calculate the probability of seeing $M$ events 
given some expected number of events $\mu$. The expected 
number of events will depend on $\alpha$, so this likelihood 
can be written as,

\begin{equation}
     P(M | \alpha, \mu) = \frac{\mu^M \left(\alpha\right) e^{-\mu(\alpha)}}{M!}.
     \label{eq-probability}
\end{equation}

A suitable model for $\mu$ must now be chosen. Assuming a 
homogeneous Poisson process, the expected number of 
events in a given interval is proportional to the duration of that interval
and the area of sky covered. This can be written as

\begin{equation}
	\mu = r_0\,\Omega\,T_{\rm obs}
\end{equation}

\noindent where $T_{\rm obs}$ is the total searchable observing time and
$r_0$ is the true rate on the sky per unit time and 
solid angle. We follow \citet{connorgbt} and tether our expected rate 
to the empirical rate of a similar survey with detections. 
This is more direct 
than the standard method of rate estimation 
which quotes an all-sky rate above 
some fixed fluence threshold and then scales accordingly with $\alpha$. 
This also eschews the need to choose a single fluence completeness value, or 
make assumptions about the distribution of pulse widths, and relaxes the 
need to account for non-uniform sensitivity over the FoV. 
The Green Bank Telescope Intensity Mapping (GBT IM) survey is a natural 
reference point, since it found the only published FRB below 1.4\,GHz and 
it overlaps with the CHIME band at 700--800\,MHz. While the GBT 
rate is consistent with the rate at 1.4\,GHz, it is more uncertain.
To account for this, we can marginalize over the low-frequency event-rate 
uncertainty from GBT. 
One could also use the more precisely determined rate from Parkes 
\citep{2016MNRAS.tmpL..49C}, but then uncertainties about scattering and 
spectral index are introduced. 
We discuss this point further in Sect.~\ref{sect-discussion}. 

We can now write down a relationship between the rate inferred 
from GBT with the number of events we expect to see at the Pathfinder.
The GBT rate is scaled in the following way

\begin{equation}
\mu_{\mathrm{PF}} = \mu_{\mathrm{GBT}} \, \frac{N^{\mathrm{PF}}_{\mathrm{days}}}{N^{\mathrm{GBT}}_{\mathrm{days}}}
\times \frac{\Omega_{\mathrm{PF}}}{\Omega_{\mathrm{GBT}}}
\times \left ( \frac{H_{\mathrm{GBT}}}{H_{\mathrm{PF}}} \right )^\alpha,
\end{equation}

\noindent where $H$ is a thermal sensitivity term given by 
the survey's bandwidth, $B$, its SEFD, $S$, and its signal-to-noise 
cut-off, $s_{\mathrm{min}}$.
Given that GBT saw one event in 27.5 days with a beamsize of 0.055\,deg$^2$, 200\,MHz
of bandwidth, and a signal-to-noise threshold of 8, we can write 
this relationship more explicitly.  
We expect the following number of events,


\begingroup\makeatletter\def\f@size{8}\check@mathfonts
\def\maketag@@@#1{\hbox{\m@th\large\normalfont#1}}%
\begin{align}
\mu_{\mathrm{PF}} = \mu_{\mathrm{GBT}}\,\frac{N^{\mathrm{PF}}_{\mathrm{days}}}{27.5} 
\times & \frac{\Omega}{0.055\,\mathrm{deg}^2} 
\left( \frac{13.25\,\mathrm{Jy}}{S_{\rm sys}} \right )^\alpha 
\left ( \frac{B}{200.0\,\mathrm{MHz}} \right )^{\alpha/2} \nonumber \\
& \,\,\, \times
\left ( \frac{\sqrt{\tau^2 + t_i^2}}{t_I} \right )^{\alpha/2}\,
\left ( \frac{8}{s_{\rm min}} \right )^\alpha.
\label{eq-mu}
\end{align}
\endgroup

\noindent In Eq.~\ref{eq-mu} $\mu_{\mathrm{PF}}$ is the expected number of events
for the incoherent Pathfinder,
and $\mu_{\mathrm{GBT}}$ is the expected number of events in 27.5 days 
of observing with GBT. The latter has a maximum-likelihood value of
1 and a 95$\%$ confidence interval of 0.25--5.57 \citep{connorgbt}. We assume GBT's SEFD to be 
26.5\,K / 2.0\,K\,Jy$^{-1}$, and we have included a DM smearing
term, which we take to be negligible in the case of Green Bank. 
We use $\mathrm{DM}=500$\,pc\,cm$^{-3}$, based on the argument in Sect.~\ref{sect-smear} 
that we are only sensitive to nearby, and therefore relatively low-DM FRBs. 
One advantage of directly extrapolating from the empirical rate of another survey
is that we do not need to compute an integral under the beam to account for direction-dependent 
sensitivity; the correction factors from the two surveys roughly divide out. However, the effect 
must be accounted for when quoting ``all-sky'' rates, especially if the telescope has 
significant sidelobes.

Using the values in Table~\ref{tab-parameters}, we can calculate the expected number 
of events, $\mu$, for 
each value of $\alpha$ and compute the probability of non-detection
with the likelihood function in Eq.~\ref{eq-probability}. 
If we were to ignore uncertainty in the rate, we would simply apply a $p$-test 
using the maximum-likelihood value in Eq.~\ref{eq-mu},
and ask what values of $\alpha$ can be ruled out with, say, 95$\%$
certainty. 
But in general, $r_0$ and $\alpha$ are degenerate \citep{Oppermann16}.
In the case of our non-detection, 
we cannot strictly differentiate between small-$\alpha$ with a low rate,
and large-$\alpha$ with a high rate. Therefore, we marginalize over 
the uncertainties in the true sky rate, similar to what is done by 
\citet{Oppermann16}. Mathematically, this is just the 
sum of likelihood curves for all rates, $r_0>0$, weighted by the 
probability density at that rate, $\mathcal{P}(r_0)$. 
We use the GBT rate posterior as $\mathcal{P}(r_0)$, and compute the following
integral,

\begin{align}
P(M=0|\,\alpha) = \int_{0}^{\infty}\!P(0|\alpha, r_0)\,\mathcal{P}(r_0)\,\mathrm{d}r_0.
\end{align}

\begin{figure}
  \centering
  \vspace{1 cm}
   \includegraphics[trim={.2in, 0in, 0.in, 0in}, width=0.48\textwidth]{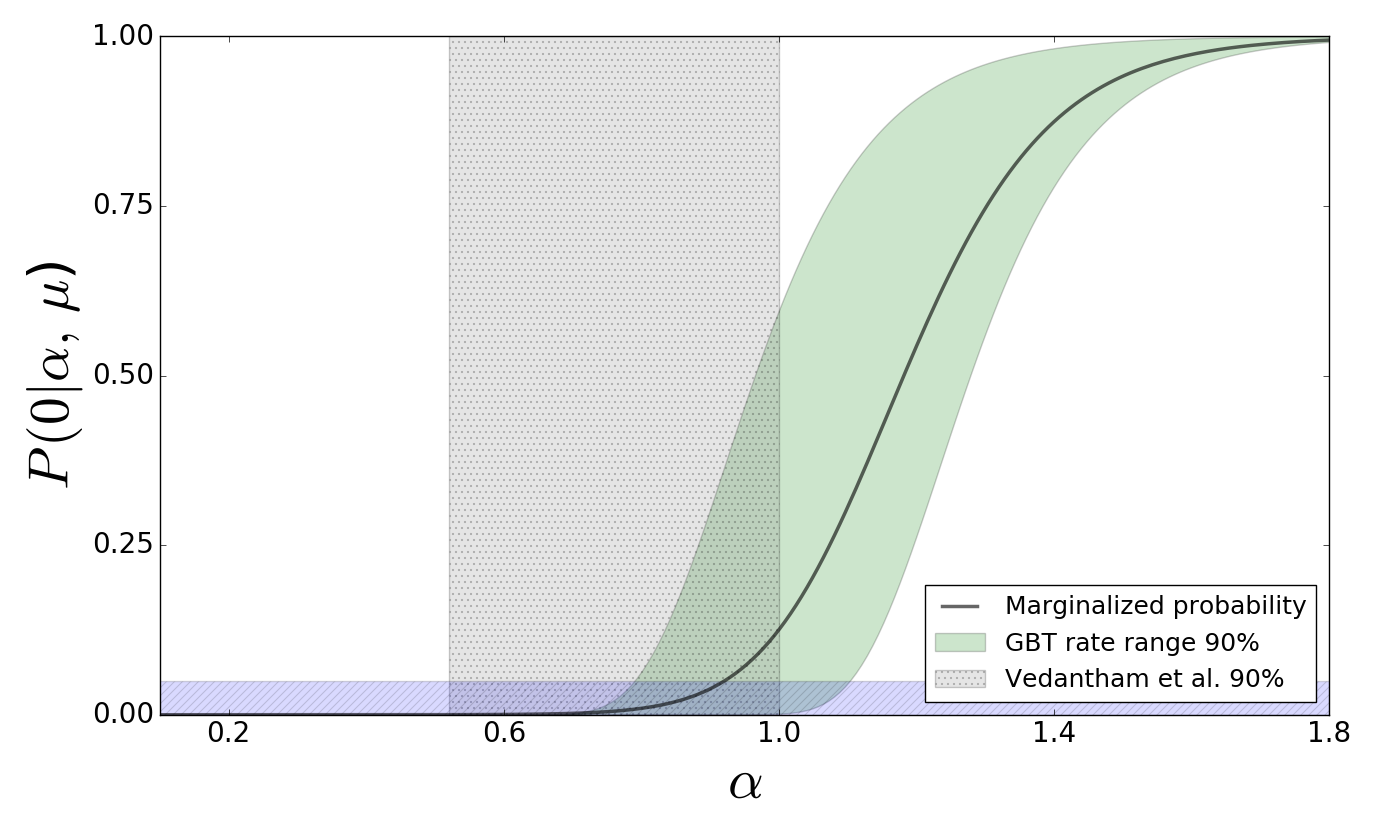}
   \caption{Probability of detecting zero events in 53 days of 
    incoherent Pathfinder data, plotted as a function of 
	the brightness distribution parameter $\alpha$. The black curve 
	shows the probability of seeing no FRBs, marginalized over 
	uncertainty in the GBT rate. The green shaded 
	region is bounded by the $95\%$ lower-limit on the GBT rate (left)
	and the $95\%$ upper-limit (right). The light blue hatched 
	region shows where the probability is less than 0.05, implying 
	that we can rule out with $95\%$ confidence $\alpha < 0.9$. 
   }
   \label{fig-constraints}
\end{figure}

This procedure produces the black curve shown in Fig.~\ref{fig-constraints}. 
The curve is equal to 0.05 at $\alpha\approx0.9$, meaning if $\alpha$ were 
smaller than 0.9, we would have expected to see one or more FRBs in 
53 days of Pathfinder data $>95\%$ of the time.
The figure also shows the non-detection 
likelihoods for a range of event rates. The green region shows 
the likelihood values for rates between 0.34-4.68 times the maximum-likelihood 
rate. 0.34 is the value above which $95\%$ of the GBT rate posterior lies, 
and 4.68 is the upper-bound on $95\%$ of the posterior. 

\section{Discussion}
\label{sect-discussion}

\subsection{Brightness-dependent $\alpha$}

The most model-independent statement we can make about our 
results is not about $\alpha$, but about the 
event rate above our sensitivity threshold, between 400--800\,MHz. 
Turning that rate upper-limit into a lower-limit 
on $\alpha$ requires some assumption about the functional form 
of the brightness distribution, and its scaling (i.e. the 
true rate on the sky). For example, we have assumed 
the distribution's shape is described by a single power-law.
But for a large enough range of brightnesses and an underlying
cosmological population, the one-parameter 
power-law assumption breaks down (see the light blue curve in Fig.~\ref{fig-standard-candles}).
Ignoring, for a moment, the Universe's star-formation history 
and considering only non-Euclidean effects, we generically expect a 
relative deficit 
of faint events. This is because FRBs at large distances will be 
diminished in energy and rate due to cosmological redshift and time dilation. 
Therefore, $\alpha$ is brightness-dependent,
flattening out for high-$z$ sources and asymptoting to 
3/2 as $z_{\rm src}$ approaches 0, with a simple mapping between brightness 
and redshift in the idealized standard-candle case. 
This phenomenon is seen in $\log\!N$--$\log\!S$ of long GRBs,
which exhibit such a continuously varying 
$\alpha$ parameter, and is nearly flat at the fluences of the faintest 
bursts. In the bright tail, the curve approaches a power-law with 
index 3/2 \citep[Fig.1]{2011MNRAS.415.3153N}. One consequence 
of this is that surveys with different sensitivities will measure 
different $\log\!N$--$\log\!S$ slopes. For example, 
the distribution of FRB signal-to-noise within a low-sensitivity survey 
may be Euclidean, even though a flatter distribution 
might be required when extrapolating to the rates of 
larger telescopes. \citet{Oppermann16} provide a framework 
for constraining $\alpha$ based on S/N distributions
as well as detection counts between surveys. 

We justified our constant-$\alpha$ assumption 
for the CHIME Pathfinder by pointing out that we were probing 
flux densities that are only a couple of orders-of-magnitude larger than 
where most FRBs have been seen ($\sim$ten times closer, on average), 
and so the effects of brightness-dependent $\alpha$ might be negligible. 
However, it is possible that $N(>\!S)$ turns over
at a flux density that is less than our threshold, and approaches $\alpha\approx1.5$
the way the brightest GRBs do. 

\subsection{Consistency with the ultra-bright rate}
The brightest event at the time of this publication, FRB 150807, 
would probably not have been detectable in our survey, 
in part because it was narrower than our sampling time and its
S/N would be reduced \citep{2016Sci...354.1249R}. Despite not having seen anything, 
we can ask if the rate of ultra-bright events implied by 150807 and the Lorimer burst
is in agreement with our results, for a given value of $\alpha$.
\citet{2016Sci...354.1249R} predict that the rate at 1.3\,GHz of FRBs above 50\,Jy\,ms  
is 190$\pm60$\,sky$^{-1}$\,day$^{-1}$. Using $\alpha=3/2$ and a minimum 
burst energy of 220\,Jy\,ms with width 5\,ms, the quoted rate predicts 
one every couple of months in our current configuration. In other words, 
our non-detection could be consistent with 190$\pm60$\,sky$^{-1}$\,day$^{-1}$ 
if those ultra-bright events are, on average, from nearby sources, and the required extrapolation 
is near-Euclidean.
On the other hand, an $\alpha=0.7$ extrapolation predicts roughly one per week, 
which is not consistent with our data. 
Of course, our non-detection is also 
consistent with the rate of \citet{2016Sci...354.1249R} having been overestimated.

\subsection{600\,MHz vs. 1.4\,GHz}

Our results are the first constraints at 600\,MHz, 
but several surveys have searched, to no avail, at lower frequencies
\citep{2015AJ....150..199T, 2014A&A...570A..60C}. 
\citet{2015MNRAS.452.1254K} searched around 140\,MHz with LOFAR 
and saw nothing, though inter-channel smearing meant they had a maximum DM
of just 320\,pc\,cm$^{-3}$. A GBT survey at 350\,MHz 
placed a $95\%$ confidence upper-limit of a few thousand detectable FRBs 
per sky per day after searching $\sim$\,80 days of data \citep{2017arXiv170107457C}.
This result is still roughly consistent with the rate at 1.4\,GHz, 
which has a lower bound around 10$^3$\,sky$^{-1}$\,day$^{-1}$ 
\citep{Oppermann16, 2016MNRAS.tmpL..49C}. Nevertheless, the uncertainty 
in FRB rate as a function of frequency is a concern for us. We 
have tried to mitigate its effects by tying the Pathfinder results to the
only published FRB survey in our band, and marginalizing 
over the rate distribution. Still, the GBT IM survey only overlaps with ours
at the top of the CHIME band, between 700--800\,MHz. If the effects of 
scattering, free-free absorption, and smearing are significantly more 
destructive in the bottom of our band, then we would have overestimated 
the effective rate and, consequently, our lower-limit on $\alpha$. 

\begin{figure}
  \centering
  \vspace{1 cm}
   \includegraphics[trim={0.in 0in 0.in 0.in}, width=0.47\textwidth]{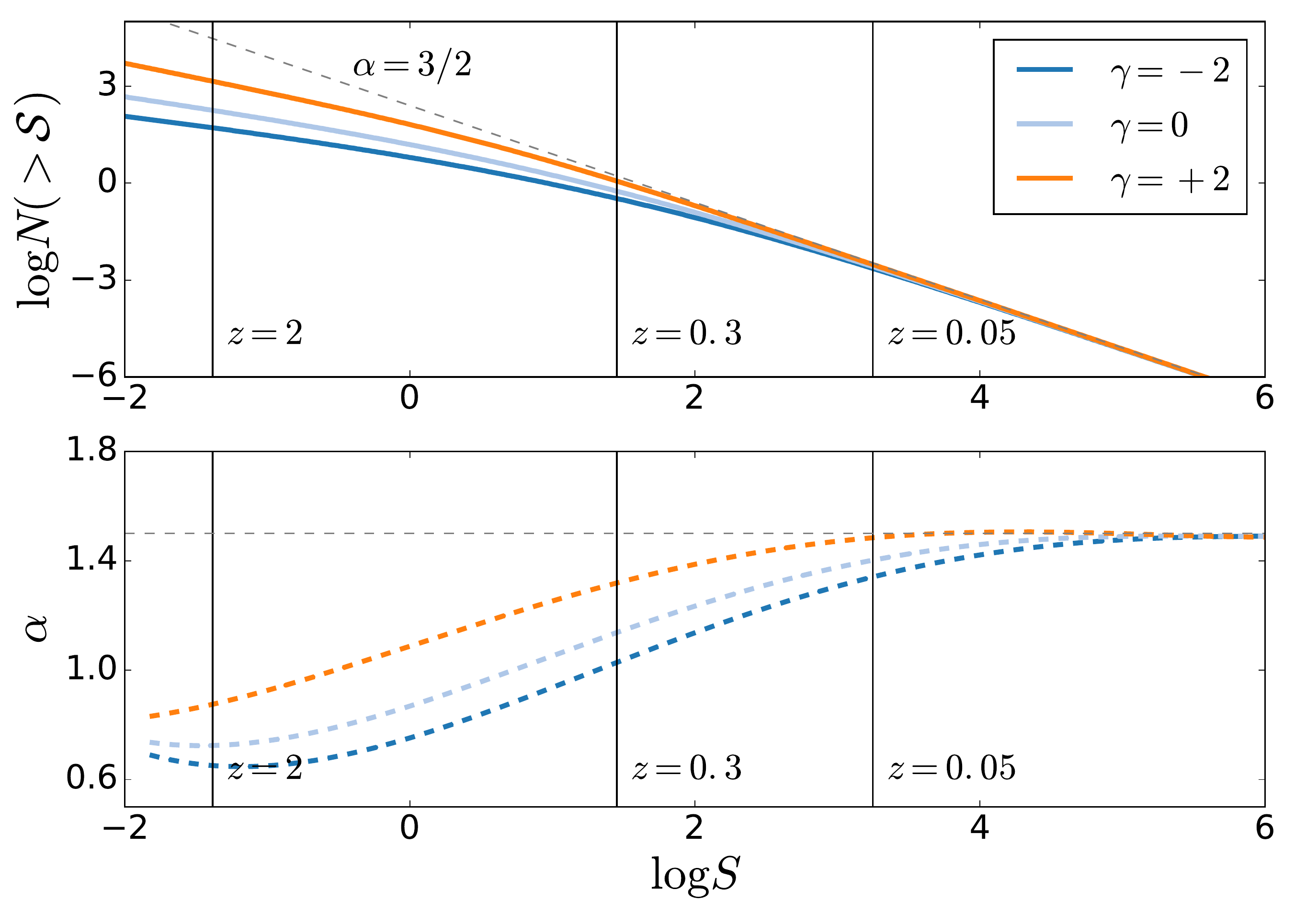}
   \caption{Simulated flux density distribution assuming standard 
   candles with varying spectral index, $\gamma$, meant to highlight how non-Euclidean
   effects and spectral index alter $\alpha$.
   The top panel shows $N(>\!S)$ as a function of $S$. The bottom 
   panel shows $\alpha$, which is the slope of these curves in logspace. 
   In the standard candle scenario, there is a one-to-one mapping 
   from fluence to redshift for a given spectral index. The vertical 
   lines show that mapping for the $\gamma=0$ case to give an idea of 
   the size of cosmological effects at each redshift. In the non-standard 
   candle case, these effects are still seen so long as the statistics of 
   intrinsic luminosity function do not evolve with distance.}
   \label{fig-standard-candles}
\end{figure}

In spite of the spectral uncertainty in rate, the brightness 
distribution's logarithmic slope, $\alpha$, should be fairly robust 
against frequency variation.
The spectral behaviour of FRBs does affect the shape of $N(>S)$ in the cosmological case, 
but at a given value of $S$, there are only special cases where 
$\alpha$ is frequency-dependent. 
If the intrinsic source luminosity is given by a power-law, 
$\mathcal{L}\propto \nu^\gamma$, then we will observe the 
part of the spectrum that has been redshifted down to our instrument's band.
Negative spectral index lowers the observed energies 
of distant FRBs as,
$\mathcal{L} \rightarrow \mathcal{L}(1+z)^\gamma$, thereby 
decreasing the number of visible distant events and flattening 
$\log N$--$\log S$. Conversely, positive $\gamma$ steepens it. These effects are 
shown in a toy-model plotted in Fig.~\ref{fig-standard-candles}.
While the curves all approach the Euclidean value of 3/2, 
there is significant spectral index dependence in $\alpha$ for 
FRBs when cosmological volumes are probed. 
If the source has non-power-law frequency behaviour (e.g. $\sim$\,GHz scintillation), 
then the source-to-source (or even pulse-to-pulse) 
variance in brightness will increase, but the ensemble distribution 
should not be affected unless there is an average tilt 
in FRB spectra.

\subsection{Implications for other surveys}
The CHIME Pathfinder's incoherent-beam survey is searching 
a limited region of FRB parameter space, namely the ultra-bright 
tail between 400--800\,MHz. Because of this large
brightness threshold, our results have few implications 
for full CHIME, which will have a flux density limit that 
is several hundred times lower than the current search, 
thanks to its coherent beams and larger collecting area. 
Therefore, the primary uncertainty in full CHIME's 
rate of detection---the deleterious effects of scattering 
and/or free-free absorption at low frequencies---remains. But as \citet{connorgbt}
showed, even if the rate between 400--700\,MHz is zero, 
CHIME's overlap with GBT IM between 700--800\,MHz indicates 
that it will see multiple bursts per day, assuming current design parameters. 
\citet{2017arXiv170107457C} also found a large event rate, 
accounting for scattering and spectral index. 
Upcoming surveys like UTMOST \citep{2016MNRAS.458..718C} 
and APERTIF \citep{leeu14} will also unite sensitivity with FoV. Therefore, their 
speed is largely $\alpha$-independent, unless they are not
operating at full capacity, e.g. during commissioning. 
Detections made in the commissioning phase, before design 
sensitivity is reached, could address our claims about brightness-dependent
$\alpha$, since those early, bright bursts, may have a Euclidean distribution.

The non-detection does, however, have implications for other
lower-sensitivity surveys. The Deep Synoptic Array (DSA) initially will consist 
of ten 5-m dishes combined incoherently, in the hopes of detecting 
ultra-bright FRBs. Saving to disk buffered voltage data could 
achieve $\sim$\,arcsecond localization, allowing for a very high-impact survey 
on a moderate budget. However, if $\alpha$ is not significantly smaller than 1.5, 
then that survey may not detect an event for many months. Extrapolating 
from the Parkes Multi Beam rate of one event every couple of weeks, we estimate that
the DSA would have to wait of order a year per FRB if $\alpha\approx1.1$.
However, given the importance of localization, a scaled-up DSA with 
more dishes could prove highly valuable. In a similar vein, the 
Australian Square Kilometre Array Pathfinder's (ASKAP) small dishes 
and phased-array feeds will effect large sky coverage with long baselines, 
potentially providing regular localization \citep{2010PASA...27..272M}.

\section{Conclusions}
\label{sect-conclusions}

We have performed a shallow, wide-field FRB survey using the CHIME Pathfinder.
This was motivated by recent assertions about the 
flatness of the brightness distribution of FRBs by \citet{vedantham2016}, who showed 
that $\alpha$ may be less than 1. 
If this were the case, the incoherent-beam Pathfinder search would be a highly competitive 
survey, potentially detecting multiple events per week. And if $\alpha$ were not quite 
so low, our search could demonstrate this with relatively little time on sky. 
We took 52.85 days of data, amassing an enormous exposure,
with $\sim$\,2.4\,$\times\,10^5$\,deg$^2$\,hrs. 
These data were searched using tree-dedispersion software 
that was used to discover FRB 110523 \citep{masui-2015b}. 
Thousands of triggers above our S/N threshold of 10 were produced, 
including daily Crab GPs and B0329+54 pulses, 
but no FRBs were found. By not detecting anything FRB signatures
we are able to rule out $\alpha<0.9$ with 95$\%$ confidence, 
using the GBT 700--900\,MHz rate and
assuming the single-index power-law approximation holds 
into our flux sensitivity. 
This constrains the number of events brighter than 
$\sim$\,220$\sqrt{(\tau/\rm ms)}$ Jy\,ms for $\tau$ between 
1.3 and 100\,ms to 
fewer than $\sim$\,13\,sky$^{-1}$\,day$^{-1}$. We quote 
our upper-limit in this way because surveys have a single 
signal-to-noise threshold, but in fluence space this cut-off  
is a curve that depends on pulse width.
The sub-arcsecond localization of FRB 121102 has shown that 
FRBs are distant enough that non-Euclidean effects 
ought to be significant. Still, its considerable local dispersion 
means that the IGM contribution is only about half
of 121102's extragalactic DM \citep{2017ApJ...834L...7T}. 
If local dispersion were a generic property of 
FRBs, then the volumes that modern surveys are sensitive to would shrink 
and deviation from $\alpha=3/2$ should be decreased, other things 
being held equal. 

As the lower-limit on $\alpha$ increases, the incoherent-beam 
Pathfinder search experiences diminishing returns in its ability to
constrain. For example, with just 5 days on sky, 
$\alpha\lesssim0.6$ can be ruled out by a non-detection with 
95$\%$ confidence. As we have shown, $\sim$\,53 days
sets a lower-bound of 0.9, but zero events in an entire year on sky 
can only rule out $\alpha\lesssim1.15$. For this reason, 
if we choose to run the incoherent-beam Pathfinder search indefinitely, 
the best strategy is to increase its sensitivity. This would mean 
investigating further the larger-than-expected noise fluctuations 
on short time-scales, perhaps mitigating it with baseband RFI 
removal. The null result suggests similar wide-field 
low-sensitivity surveys may not be highly competitive, 
but has little implication for wide-field deep surveys 
like full CHIME, APERTIF, and UTMOST. 

\begin{acknowledgements}
We are very grateful for the warm reception and skillful
help we have received from the staff of the Dominion Radio
Astrophysical Observatory, which is operated by the National
Research Council of Canada. The CHIME Pathfinder is funded by grants from 
the Natural Sciences and Engineering Research Council (NSERC), 
and by the Canada Foundation for Innovation (CFI).
LC acknowledges that the research leading to these results has received 
funding from the European Research Council under the 
European Union's Seventh Framework Programme 
(FP/2007-2013) / ERC Grant Agreement n. 617199. We also 
thank Manisha Caleb and Casey Law for useful discussions.
\end{acknowledgements}

\bibliography{pathfinder_aas}
\bibliographystyle{aastex61}

\end{document}